\documentclass{elsart}
\usepackage{natbib}
\usepackage{graphicx,amssymb}
\usepackage{filecontents}
\usepackage{longtable}
\usepackage[utf8]{inputenc}
\usepackage[english]{babel}
\usepackage{caption}
\usepackage{pdfpages}
\usepackage{epstopdf}
\usepackage[graphicx]{realboxes}
\usepackage{verbatim}
\usepackage{array}
\usepackage{longtable}
\usepackage{pdflscape}
\usepackage{subcaption}

\makeatletter
\def\elsartstyle{%
	\def\normalsize{\@setfontsize\normalsize\@xiipt{14.5}}
	\def\small{\@setfontsize\small\@xipt{13.6}}
	\let\footnotesize=\small
	\def\large{\@setfontsize\large\@xivpt{18}}
	\def\Large{\@setfontsize\Large\@xviipt{22}}
	\skip\@mpfootins = 18\p@ \@plus 2\p@
	\normalsize
}
\makeatother

\def\astrobj#1{#1}
\def\url#1{{\ttfamily\def\/{/\discretionary{}{}{}}#1}}

\begin{document}
\bibliographystyle{plainnat}

\begin{frontmatter}
\title{Absolute Parameters of the Southern Detached Eclipsing Binary DG Mic}

\author{Derya S\"{u}rgit$^{1,2}$, M\"{u}cahit Kuz$^{3}$, Volkan Bak{\i}\c{s}$^{4}$}
\author{Chris Engelbrecht$^{5}$ and Fred Marang$^{5}$}

\address{$^{1}$Astrophysics Research Center and Ulup{\i}nar Observatory, \c{C}anakkale Onsekiz Mart University, TR-17100, \c{C}anakkale, T\"{u}rkiye\\
$^{2}$Department of Space Sciences and Technologies, Faculty of Science, \c{C}anakkale Onsekiz Mart University,\\ Terzio\u{g}lu
Kamp\"{u}s\"{u}, TR-17100, \c{C}anakkale, T\"{u}rkiye \\
$^{3}$ School of Graduate Studies, \c{C}anakkale Onsekiz Mart University, TR-17100, \c{C}anakkale, T\"{u}rkiye\\
$^{4}$ Faculty of Sciences, Department of Space Sciences and Technologies, Akdeniz University, TR-07058, Antalya, T\"{u}rkiye\\
$^{5}$ Department of Physics, University of Johannesburg, PO Box 524, Auckland Park 2006, South Africa\\
}

\thanks[email]{E-mail: dsurgit@comu.edu.tr}

\begin{abstract}
As part of an ongoing programme of observing detached eclipsing binary stars in the southern sky, we present the first analysis of spectroscopic observations of the Algol-type binary system DG Mic. A spectroscopic analysis of mid-resolution spectra allowed us to constrain the effective temperature of the primary component and to test the consistency of the system parameters with its spectral energy distribution (SED). Combined solutions of mid-resolution spectra and TESS, ASAS and WASP light curves imply a system of two almost identical components ($q$ = 0.99) in circular orbits. Our final model shows that the system is a detached binary star. The masses and radii of the primary and secondary components of DG Mic were derived to be 1.65($\pm$0.12) M$_\odot$, 1.64($\pm$0.18) M$_\odot$ and 1.63($\pm$0.10) R$_\odot$, 1.91($\pm$0.13) R$_\odot$, respectively. According to Geneva evolution models, both components of the system are main-sequence stars and their age is approximately 713 Myr.

\end{abstract}

\end{frontmatter}

\newpage

\section{Introduction}
\label{Section1}

The physical parameters of eclipsing binary (EB) stars may be derived with considerable accuracy and precision by combining photometric and spectroscopic observations of EB systems. The masses, radii, and surface temperatures of the components of EB systems may be calculated by essentially using Kepler's laws of motion and radiation thermodynamics. In addition, the age of an EB system may be estimated by comparing the physical parameters obtained from the spectroscopic and photometric analysis with theoretical evolutionary tracks. Because the majority of stars appear to be members of binary or multiple systems, the knowledge gained from EB studies is vital to understanding the evolution of real populations in the Galaxy as a whole. However, the number of EB systems for which the physical parameters of the component stars have been determined with high precision is still small. The number of analyses of \lq\lq{twin systems}\rq\rq (for which the mass ratio is practically equal to 1) is even  smaller \citep{Bulut_2021}, \citep{Yucel_2022}. The number of well-analysed EB systems needs to be increased substantially before statistically significant conclusions about stellar evolution may be drawn. With this study, we extend the database of well-analysed EB systems in general, and of twin systems in particular. 

DG Mic (CPD-43 9403 = HD 201964 = Gaia DR2 6580038335837487744) was first listed as a binary system, of spectral type A0, in the catalogue of \citet{Strohmeier_etal_1965}. The spectral type of the system was designated as A2  by \citet{Drilling_and_Philip_1970}, with a Johnson $V$ magnitude of 8.4. The spectral type was listed as A2mA7/8-A8/9 in the catalogue of \citet{Houk_1978}. Later, it was included in a catalogue of 80 eclipsing binary stars by \citet{Otero_2004} and also in the catalogue of variable stars of \citet{Kazarovets_etal_2008}. DG Mic was also recorded as a variable star in a catalogue of Ap and Am type stars by \citet{Renson_etal_1991}. According to the ASAS 3 catalogue \citep{Pojmanski_2002}, DG Mic is an Algol-type variable star with spectral type A2mA7/8-A8/9, $V$ magnitude varying between 8.38 and 8.84 mag, and a photometric period of 2.69592 days. The system has been included in many catalogue studies. We could not find any spectroscopic study of the system recorded in the literature. For this reason, it was decided to observe the system spectroscopically and analyze it in more detail by combining our spectroscopic observations with existing photometric light curves from various databases.

The paper is organised as follows: \S\ref{Section2} describes the observations and data reduction; \S\ref{Section3} outlines the radial‑velocity analysis and determination of the mass ratio; \S\ref{Section4} presents the spectroscopic analysis for the temperature determination and photometric modelling; finally, \S\ref{Section5} discusses the derivation of the absolute parameters, SED fitting, distance estimation (including a comparison with Gaia data), and the evolutionary scenario of the system.

\section{Spectroscopic observations and data reduction}
\label{Section2}

The spectroscopic observations of DG Mic were obtained with the newly improved Cassegrain (SpUpNIC) instrument mounted at the Cassegrain focus of the 1.9-m telescope at the South African Astronomical Observatory (SAAO) (see \citet{Crause_etal_2016,Crause_etal_2019}). We selected a grating of the spectrograph which has 1200 lines/mm, a wavelength coverage of 400 –- 525 nm, a blaze peak at 510 nm, and a resolution of 0.06 nm (corresponding to a resolution of approximately 40 km ${\rm s}^{-1}$ in radial velocity (RV) and an approximate resolving power $R = 8000$  at the wavelength of the H$\beta$ line). A total of 39 spectra of DG Mic were obtained during the 2020 observing season (26 in August and 13 in October). Some of the spectra were excluded from the analysis because of excessive noise identified during reductions. The spectroscopic observations of DG Mic were accompanied by contemporaneous observations of radial velocity (RV) standards HD 693 (F8V, $V_r = 14.81$ km ${\rm s}^{-1}$) and HR 6031 (A1V, $V_r = -5.10$ km ${\rm s}^{-1}$), which have spectral types close to that of DG Mic. During all spectroscopic observations of DG Mic, the average exposure time was between 1000 and 1300 s, depending on the brightness of the system and weather conditions at the time of observation. Cu/Ar arc spectra were taken as comparison spectra before and after each stellar image. A set of quartz-iodine lamp images was also taken every night for flat-field calibrations. Standard IRAF procedures were used for the reduction and calibration of the spectroscopic data.


\section{\textbf{Radial velocity measurement and orbital parameters}}
\label{Section3}

\begin{figure}[ht]
        \centering
\includegraphics[width=135mm]{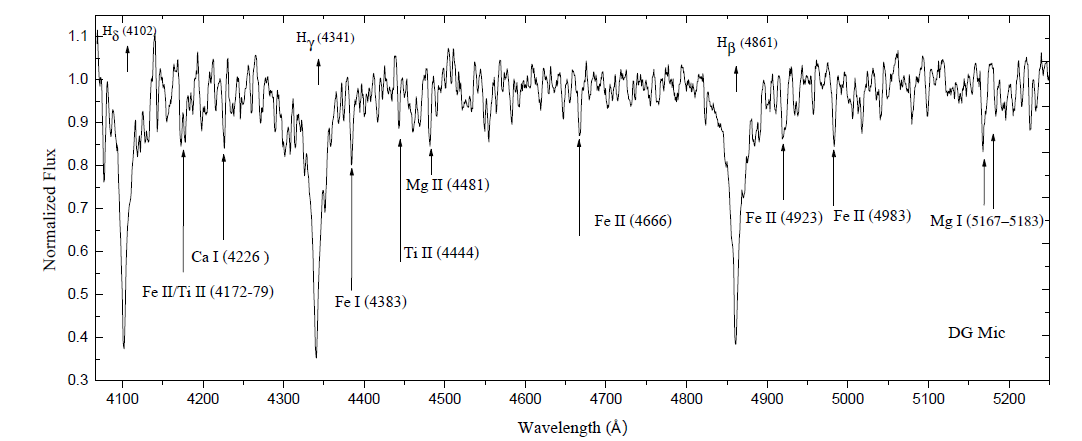}
\caption{Sample spectrum of DG Mic for orbital phase 0.5. The hydrogen Balmer lines, Mg II (4481) line, and many metallic lines are visible in the spectrum.}
\label{fig1.png}
\end{figure}

\begin{figure}[ht]
        \centering
\includegraphics[width=130mm]{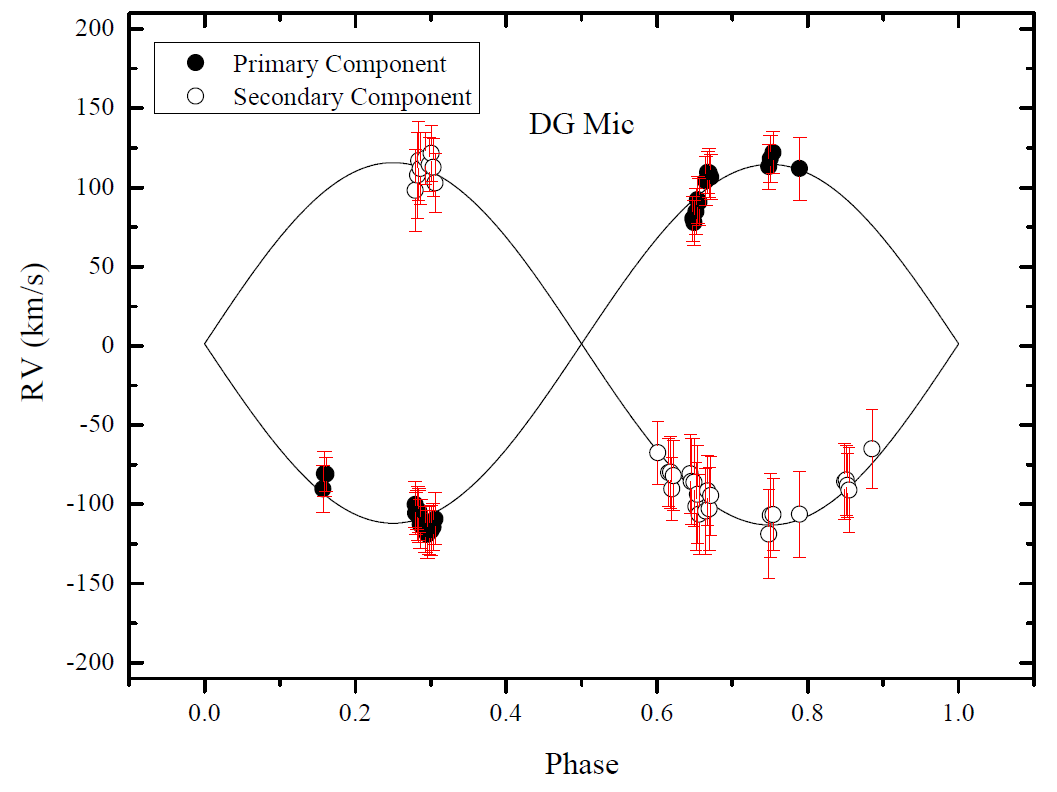}
\caption{Radial velocities of DG Mic obtained using the cross-correlation method (open and filled circles) and the best theoretical fit (solid line).}
\label{fig2.png}
\end{figure}

The radial velocities of the components of DG Mic were calculated using the cross-correlation method with the IRAF package {\sc Fxcor} \citep{Tonry_Davis_1979, Popper_Jeong_1994}. The Mg II (4481) line, which was the most prominent line after the hydrogen Balmer lines (see Fig. \ref{fig1.png}) was used to calculate radial velocities for DG Mic. In contrast, the metallic lines (Ti II 4444, Mg II 4481, Fe II / Ti II 4172-79, and Fe II (4666, 4923, 4983)) in the spectral region 4440–4600 {\AA} (see Fig. \ref{fig1.png}) were used for the cross-correlation measurements of DG Mic. The radial velocity values (with standard errors) of the components of DG Mic, calculated using the cross-correlation method, are listed in Table \ref{TableA1}. Because of the low resolution of the spectroscopic data, the radial velocity values could not be measured for both components at some phases. The radial velocity values for only one component were retrieved in these instances. The program ELEMDR77 (http://www.astro.sk/~pribulla/soft.html) was used to derive orbital parameters of DG Mic, using the calculated RVs of the components. For the determination of the spectroscopic orbital parameters of DG Mic, the orbital period was fixed at the value of 2.69592 days listed in the ASAS database \citep{Pojmanski_2002}. The orbit of the binary system was assumed to be circular. The velocity amplitudes ($K_{1}$ and $K_{2}$) of the components and the conjunction time ($T_{0}$) were taken as free parameters. The best-fitting orbital elements of DG Mic are listed in Table \ref{table1}, and the corresponding best theoretical fits to the radial velocity curves are shown in Fig. \ref{fig2.png}.

\begin{table}
\begin{center}
\caption{The orbital parameters of DG Mic.} \label{table1}
\begin{tabular}{lc}
\hline
Parameter	& Value\\
\hline
$P_{orb}$ (days)	 & 2.695920(fixed) \\
$T_{o}$ (HJD)		 & 52104.7293${\pm}$0.0008  \\
$V_{\gamma}$ (km/s)   & 1.33${\pm}$1.02 \\
$K_{1}$ (km/s)    & 113.25${\pm}$1.56 \\
$K_{2}$ (km/s)    & 114.38${\pm}$1.62 \\
$q_{sp}=M_2/M_1$            & 0.990${\pm}$0.044 \\
$A_{1}{\sin}i$ (AU) & 0.0281${\pm}$0.0004 \\
$A_{2}{\sin}i$ (AU) & 0.0283${\pm}$0.0004 \\
$M_{1}{\sin}^{3}i$ (M$_{\odot}$) & 1.65${\pm}$0.05 \\
$M_{2}{\sin}^{3}i$ (M$_{\odot}$) & 1.64${\pm}$0.05 \\
\hline
\end{tabular}
\end{center}
\end{table}

\section{Photometric Analysis}
\label{Section4}

Photometric observations of DG Mic have been made by various ground-based and/or space-based surveys. We used the light curve data from ASAS-3 \citep{Pojmanski_2002}, the Transiting Exoplanet Survey Satellite (TESS) \citep{Ricker15}, and the Wide Angle Search for Planets (WASP) \citep{Butters10} for our photometric analysis of DG Mic. DG Mic was observed by TESS with a cadence of 1800 seconds in sector 1 (2018 July 25th), with a cadence of 600 seconds in sectors 27 and 28 (2020 July 5th, 2020 July 31st), and with a cadence of 200 seconds in sector 68 (2023 July 29th). The  TESS data of DG Mic were accessed via the Mikulski Archive for Space Telescopes (MAST, https://archive.stsci.edu/) database.
The light curves of DG Mic in the TESS, ASAS, and WASP databases were analysed using the Wilson-Devinney (WD) code \citep{Wilson71} combined with the Monte Carlo (MC) search method \citep{Zola_etal_2004, Zola_etal_2010}. In this method, the binary system’s light curve is analysed by accounting for ellipticity, tidal interactions, and the gravitational equipotential surfaces of the stars. The method performs hundreds of thousands of iterations in the solution space until it produces the best theoretical curve fit to the observational light curve(s).

The WD+MC  method consists of three steps. Accordingly, (i) the TESS, ASAS and WASP light curves were resolved by using the parameters and input ranges defined below -- we used the mass ratio determined from the RV solution as an input value in this step; (ii) the RV curve was solved taking into account the tidal and rotational effects determined in the first step (in the form of elliptic and proximity effects) -- in this process, a corrected value of the mass ratio $q_{corr}$ was produced; (iii) in the final step, the actions performed in step (i) were repeated using $q_{corr}$ as the new input value for the mass ratio.

\subsection{Temperature Determination}

Before starting the light curve analysis, it is essential to obtain a reliable estimate of the effective temperature of at least one component in the system. The temperature can be most accurately constrained from spectroscopic data. In our spectroscopic dataset, the spectrum obtained at orbital phase 0.5---when the primary component is exactly in front of the companion---directly provides information about the temperature of the primary.

To model this spectrum, we constructed a synthetic spectral grid based on \textsc{Atlas9} models with new opacity distribution functions (ODFs) \citep{Castelli2003}, covering:
\begin{itemize}
    \item a temperature range of 7000--8000~K with a step of 100~K,
    \item a metallicity range of $-0.5$ to $+0.5$~dex with a step of 0.1~dex,
    \item a projected rotational velocity ($v \sin i$) range of 50--150~km\,s$^{-1}$ with a step of 10~km\,s$^{-1}$.
\end{itemize}

The surface gravity was fixed at $\log g = 4.24$ (cgs), and the microturbulence velocity at $\xi = 2$~km\,s$^{-1}$. 
For the observed spectrum, we selected the wavelength range 4750--5200~\AA, which includes the H$_\beta$ line and relatively strong metallic lines.

\begin{figure}[ht]
        \centering
\includegraphics[width=130mm]{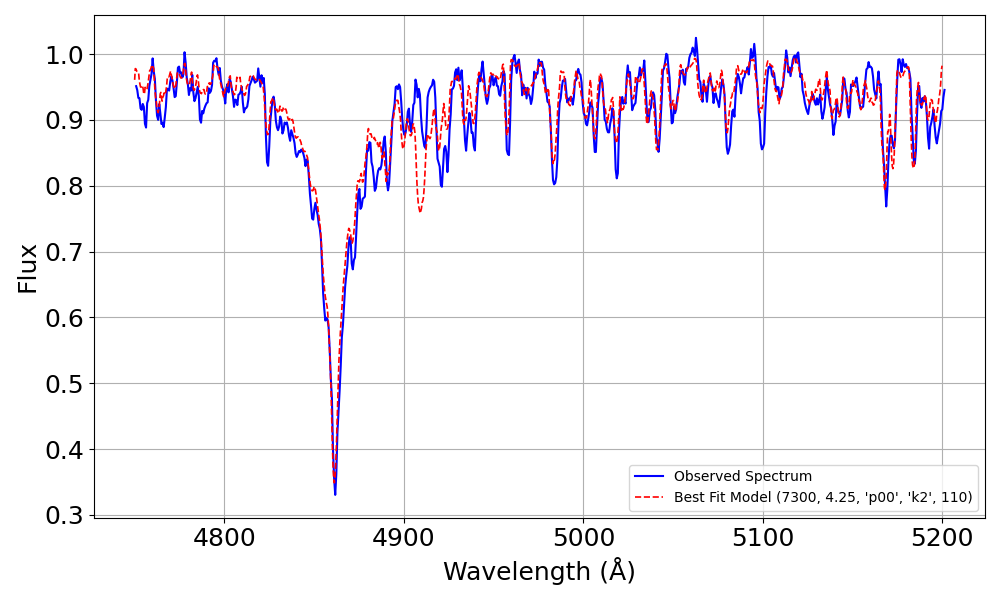}
\caption{Observed spectrum of the primary component at orbital phase 0.5 (blue solid line) together with the best-fitting synthetic spectrum (red dashed line) generated from the \textsc{Atlas9} models \citep{Castelli2003}. 
The best-fit parameters are $T_{\rm eff} = 7300 \pm 150$~K, 
$v \sin i = 110 \pm 10$~km\,s$^{-1}$, and [m/H] $= 0.0 \pm 0.15$~dex. The comparison was performed in the wavelength range 4750--5200~\AA, covering H$_\beta$ and strong metallic lines.}
\label{fig:sp_model}
\end{figure}

Within the constructed grid, the best-fitting model (minimum $\chi^2$) corresponds to $T_{\rm eff} = 7300$~K, $v \sin i = 110$~km\,s$^{-1}$, and [m/H] = 0.0~dex. The observed spectrum together with the best-fitting model is shown in Figure~\ref{fig:sp_model}.

\begin{figure}[ht]
  \centering
  \begin{subfigure}[t]{0.48\textwidth}
    \includegraphics[width=\textwidth]{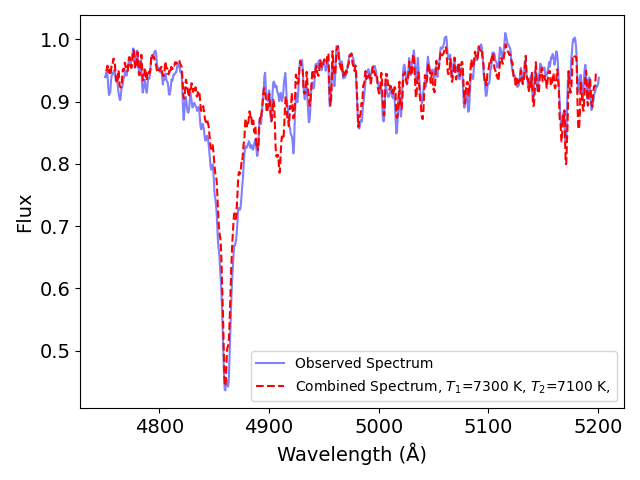}
  \end{subfigure}
  \hfill
  \begin{subfigure}[t]{0.48\textwidth}
    \includegraphics[width=\textwidth]{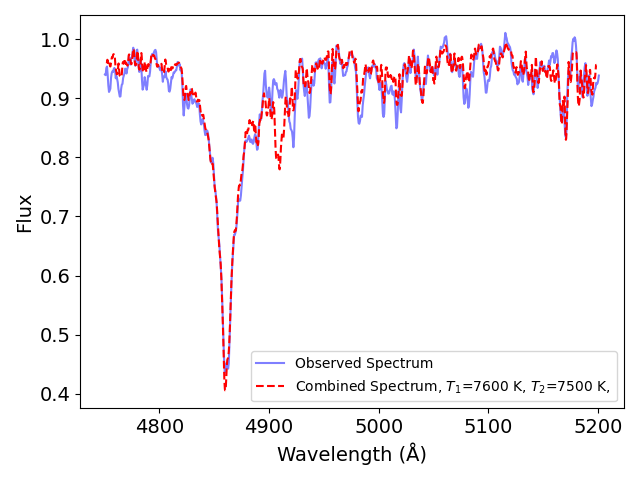}
    \label{fig:spectrum2}
  \end{subfigure}
  \caption{Synthetic spectrum models for spectra taken at phase 0.25. \textit{Left:} Model with relatively lower temperature components fits better to metallic lines. \textit{Right:} Model with relatively higher temperatures fits better to $H\beta$ line.}
      \label{fig:composite_model}
\end{figure}

To verify the accuracy of our spectral model at phase 0.5, we compared the observed spectrum with synthetic spectra constructed at phase 0.25. In this process, both stellar components were shifted along the wavelength axis according to their Doppler shifts, and their flux contributions were weighted using the light ratios derived from the light curve. As shown in Figure~\ref{fig:composite_model}, although an overall agreement is achieved, the H$\beta$ line favors a slightly hotter temperature combination ($T_{1}=7600$~K, $T_{2}=7500$~K), whereas the metal lines are better reproduced by a cooler combination ($T_{1}=7300$~K, $T_{2}=7100$~K). This behavior is commonly observed in Am-type stars. Since our model at phase 0.5 yields a smaller $\chi^{2}$ value for $T_{1}=7300$~K, we adopted this temperature for the primary component.

The uncertainties in the derived parameters were estimated from the $\chi^2$ distribution around the best-fit model, adopting $\Delta\chi^2 = 1$ as the 1$\sigma$ confidence interval. Taking into account the spectral signal-to-noise ratio and the finite grid steps (100~K in $T_{\rm eff}$, 0.1~dex in [m/H], and 10~km\,s$^{-1}$ in $v \sin i$), we adopted conservative uncertainties of $\pm 150$~K in $T_{\rm eff}$, 
$\pm 0.15$~dex in [m/H], and $\pm 10$~km\,s$^{-1}$ in $v \sin i$.

\subsection{Light Curve Modeling}
\begin{figure}
        \centering
\includegraphics[width=130mm]{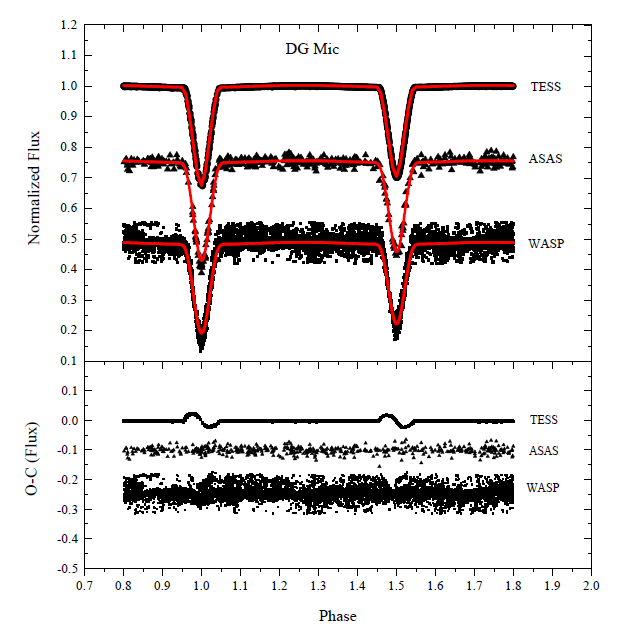} \\
\includegraphics[width=80mm]{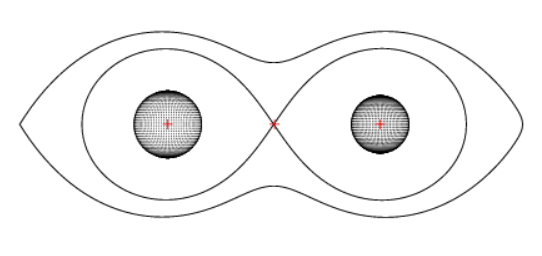}
\caption{{\color{black} Upper panel: TESS, ASAS and WASP light curves of DG Mic and the best theoretical fits; Lower panel: Roche geometry of the DG Mic system.}}
\label{fig3}
\end{figure}

After estimating the effective temperature of the primary component, we searched for the best-fitting solution within a selected range for each free parameter using the Monte Carlo (MC) method.
The selected ranges were $70^\circ < i < 90^\circ$ for the orbital inclination of the system, $T_2$=6500 – 7300 K for the effective temperature of the secondary component, $\Omega _1$=6.0–10.0 and $\Omega _2$=5.0-7.7 for the dimensionless surface potentials of the components, and $-0.01 < \Delta \phi <0.01$ for the phase shift. The fractional luminosity of the primary component ($L_1/(L_1+L_2)$) and  the possibility of a third body contributing to the total light was evaluated by taking ($l_3$) into account as a free parameter in the analysis.  During the solutions, the spectral mass ratio ($q=M_2/M_1$) was fixed at the constant value of 0.990 ($\pm0.044$) determined from the radial velocity solution. The parameters that were taken as constants during the solutions included the effective temperature of the primary component ($T_1$), the square-root limb-darkening law taken from \citet{Claret_and_Bloemen_2011} and \citet{Claret_etal_2013} (depending on the filter wavelengths used and the effective temperatures of the components), and the bolometric gravity darkening coefficients of the constituents (we used the value of 1.0 for radiative atmospheres from \citet{von_Zeipel_1924} when $T>7200$ K, and the value of 0.32 for convective atmospheres from \citet{Lucy_1967} when $T<7200$ K). For bolometric albedo coefficients of the components, we used 1.0 for radiative atmospheres and 0.5 for convective atmospheres, from \citet{rucinski1969}. The orbits were assumed to be circular ($e=0$) during all iterations. The system was solved with MODE 2 for a detached binary configuration model. The final model for DG Mic resulting from the analysis is listed in Table \ref{table2}. 
Given that the TESS data provide higher sensitivity compared to the ASAS and WASP observations, all subsequent calculations were performed using the TESS results. The best theoretical fits relative to the TESS, ASAS, and WASP light curves, repectively, are shown in the upper panel of Fig. \ref{fig3}. The larger scatter in the residuals of the TESS data around eclipse phases may be attributed to the smaller phase bins used at eclipse phases. Comparable behavior has also been reported in previous studies, such as those by \citet{Maceroni_etal_2014} for KIC 3858884 and \citet{Erdem_etal_2024} for SY Phe. According to \citet{Maceroni_etal_2014}, this scatter arises from the representation of the stellar surface in PHOEBE (and WD), in which the edge of the first tile along each meridian produces a seam along the central line oriented toward the companion star.

The fill-out factor f utilized in Binary Maker 3 is a modification of the formulation by \citet{Lucy_Wilson_1979}. Following the determination of the system’s mass ratio, it enables the specification of equipotential surfaces for detached binary systems as well as for contact and over-contact binary systems. The fill-out parameter characterizes the degree of deviation of a binary component from contact, indicating whether the system is in an under-contact (detached) or over-contact configuration. For under-contact configurations, the fill-out factor is defined as the normalized difference between the surface potential of the system and the inner critical potential, minus 1. The fill-out factor f for detached stars will lie in the range from -1 to 0. From our analysis, DG Mic was identified as a detached binary system. Accordingly, the fill-out parameter f was derived using Equation (1).

\begin{equation}
f\equiv\frac{{\Omega}_{inner}}{\Omega}-1    
\label{eq1}
\end{equation}

Consequently, the fill-out parameters f calculated for the components of DG Mic were -0.540 and -0.472, respectively. Our final solution for DG Mic using TESS data indicates that it is a detached eclipsing binary system where the primary and secondary components fill 46\% and 53\% of their Roche limiting lobes, respectively. The Roche geometry of the system, obtained with Binary Maker \citep{Bradstreet02}, is shown in the lower panel of Fig. \ref{fig3}.


\begin{table}
\begin{center}
\caption{Parameters of final model of TESS, ASAS and WASP light curves of DG Mic with WD+MC method.}
\label{table2}
\begin{tabular}{lccc}
\hline
Parameters & TESS & ASAS & WASP  \\
\hline
$i$ ($^\circ$)    & 85.19($\pm$0.25) & 86.51($\pm$0.30)   &    84.92($\pm$0.23)   \\
$\phi$                     & -0.0011($\pm$0.0001)  & 0.0005($\pm$0.0002) & 0.0003($\pm$0.0002)     \\
$T_1$ (K)         & 7300 (fixed) & 7300 (fixed) & 7300 (fixed)                \\
$T_2$ (K)         & 7025($\pm$204)  & 7090($\pm$219)  & 7091($\pm$220)    \\
$\Omega_1$        & 8.397($\pm$0.160)  & 8.468($\pm$0.191)  & 8.500($\pm$0.170)  \\
$\Omega_2$        & 7.292($\pm$0.089) & 7.216($\pm$0.102)  &  7.300($\pm$0.100) \\
$r_1$ (volume)    & 0.135($\pm$0.006) & 0.134($\pm$0.008)  & 0.133($\pm$0.010) \\
$r_2$ (volume)    & 0.158($\pm$0.008) & 0.160($\pm$0.009)   &  0.158($\pm$0.013) \\
$L_1/(L_1+L_2)$ (TESS)       & 0.422($\pm$0.019) &  &     \\
$L_1/(L_1+L_2)$ (ASAS)       &  &  0.372($\pm$0.021)  &    \\
$L_1/(L_1+L_2)$ (WASP)       &      &          &               0.398($\pm$0.025)    \\
$l_3$ (TESS)            &   0.056($\pm$0.018)    &    &        \\
$l_3$ (ASAS)            &  & 0.155($\pm$0.026)   &        \\
$l_3$ (WASP)          &    &                     &   0.086($\pm$0.023)     \\
\hline
\end{tabular}
\end{center}
\end{table}

\section{Results and Discussions}
\label{Section5}

The RV curves of the components of DG Mic were obtained from our SAAO spectra by using the cross-correlation method. We determined the spectroscopic mass ratio of the system as 0.990 ($\pm0.044$). The light curves of DG Mic in the TESS, ASAS, and WASP databases were analyzed using the Wilson-Devinney method combined with the Monte Carlo search method.

\subsection{Astrophysical parameters and the SED model}

{Table~\ref{table2} lists the final solutions for DG Mic, together with the formal 1‑$\sigma$ errors derived from the WD Monte‑Carlo runs.  The effective temperature of the primary component was determined directly from the medium‑resolution spectra using the line‑depth ratios and Balmer‑line profile fitting (see \S3). This spectroscopic approach yields $T_1=$7300$\pm$150 K, which we adopted as a fixed value in the WD analysis.  The secondary temperature, $T_2=$7025$\pm$204 K, follows from the light‑curve solution and is fully consistent with the spectral type inferred from the spectra (A7/8 V).

Using the solar values of $T_{eff}=5771.8(\pm0.7)$ K, $M_{bol}=4.7554(\pm0.0004)$ mag, $BC=-0.107(\pm0.020)$ mag, and $g=27423.2(\pm7.9)$ cm/s$^2$ \citep{Pecaut_and_Mamajek_2013}, we calculated the bolometric corrections for each component from the tables of \citet{Eker_Bakis_2023}.  The absolute parameters (masses, radii, luminosities) derived in Table~\ref{table3} are fully compatible with those expected for a detached A‑type binary at this evolutionary stage. The interstellar extinction was evaluated via \textit{A$_V$} = 3.1$\times$\textit{E}(\textit{B--V}). From the distance modulus we obtain $d=$224$\pm$20 pc. The consistency between these independent distance estimates and the Gaia-derived distances (see Table 3) gives us confidence that the adopted absolute scale is robust.

With the fundamental parameters fixed, we constructed a SED for DG~Mic. The model SED was generated using the \citet{Castelli2003} atmosphere grids with the latest opacity‑distribution functions.  All available photometric points from the UV to the IR were collected within a 2 arcsec aperture around the target (VizieR). In the fitting procedure only the colour excess $E(B-V)$ was allowed to vary; all other parameters (temperatures, radii, distance, bolometric corrections) were held fixed at their WD values.  The best fit yields an almost negligible reddening, $E(B-V)=$0.00$\pm$0.01, and the synthetic SED reproduces the observed fluxes to within the quoted photometric uncertainties (see Fig.~\ref{fig:sed_model}).  This result confirms that our spectroscopic temperatures, radii, and adopted distance produce a coherent picture of the system’s energy output across the entire spectral range.

In summary, the spectroscopically derived temperature for the primary, combined with the precise spectroscopic mass ratio and well‑constrained light‑curve solution, leads to an internally consistent set of stellar parameters.  The negligible interstellar extinction inferred from the SED fitting further supports the reliability of these values. The agreement between our distance estimate and Gaia parallaxes provides an external validation of the adopted absolute scale.  Together, these results establish DG~Mic as a benchmark detached A‑type binary suitable for detailed evolutionary studies.

\begin{figure}[ht]
        \centering
\includegraphics[width=130mm]{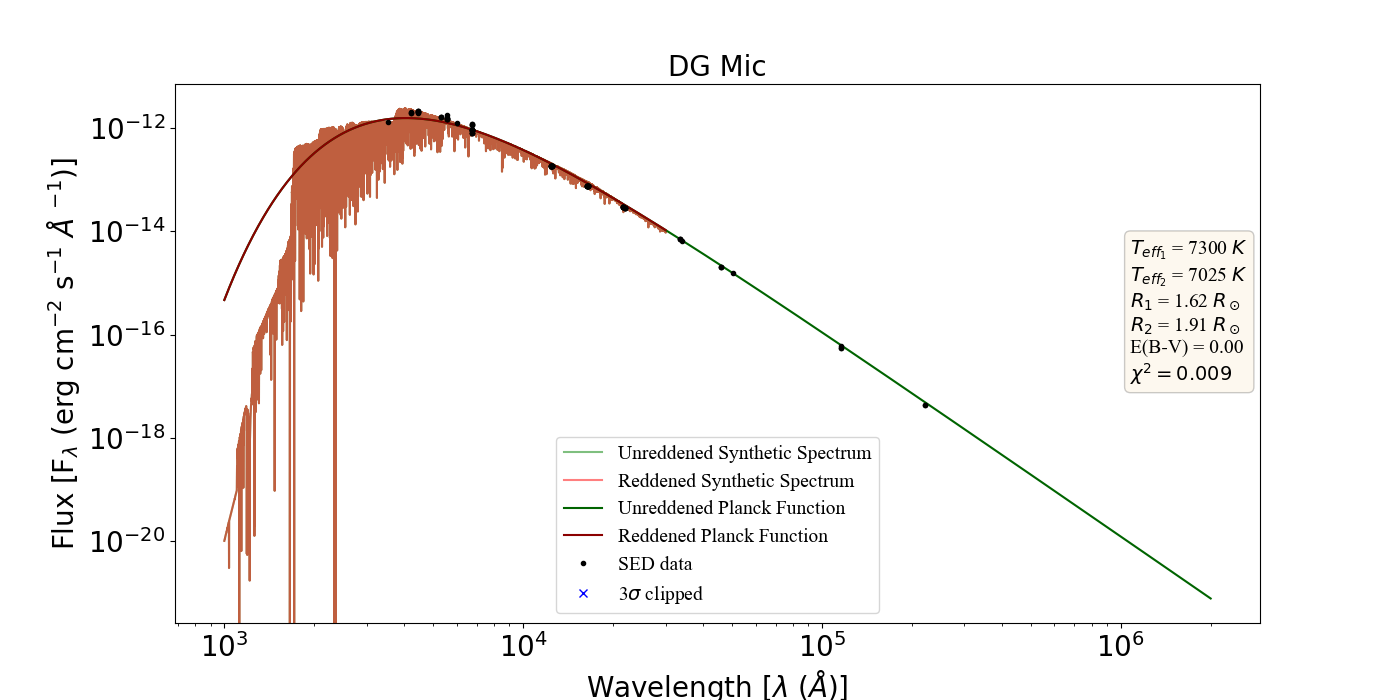}
\caption{Spectral energy distribution (SED) of DG~Mic. 
The black points represent all available photometric measurements within a 2~arcsec aperture retrieved from VizieR, covering the UV to IR range. The red solid line shows the best-fitting synthetic SED generated using the \citet{Castelli2003} model atmosphere grids with updated ODFs. During the fitting procedure, the system parameters from Table~3 were fixed, whereas only the interstellar reddening $E(B-V)$ was allowed to vary. The best fit yields $E(B-V) = 0.00 \pm 0.01$. As evident from the figure, the adopted parameters and distance are in good agreement with the available UV--IR observations.}
\label{fig:sed_model}
\end{figure}

\subsection{Evolutionary Scenario}
Geneva evolutionary models \citep{ekstrom_etal_2012} were selected to compare the measured physical parameters of DG Mic with theoretical model predictions, given chemical compositions of $Z = 0.020$ \citep{Mowlavi_etal_2012}. The isochrones of the determined ages of the components are shown in Fig. \ref{fig4} (upper panel), using the Hertzsprung–Russell (log ($T_{eff}$) - log ($L$)) diagram. The isochrones of the determined ages were checked using the mass-radius diagram in Fig. \ref{fig4} (lower panel). The calculated radii of the primary and secondary components are attained at log (age) = 8.64 and log (age) = 9.07, respectively. Referring to the plots in the HR diagram, we recognise that the isochrones for both components of DG Mic are accommodated within the standard errors by inferring a value of log (age) = 8.85, corresponding to an age of 713 Myr for the binary system. This age  corresponds to both components being main-sequence stars. The radius of the second component displayed in the lower panel of Fig. \ref{fig4} appears too high to accommodate the age of 713 Myr. A similar radius anomaly was reported as a common feature in detached eclipsing binary systems by \citet{Cruz_2022}, although that study did not extend to systems with components that had masses greater than 1 {M}$_{\odot}$, and explanations of the phenomenon have been limited to arguments that relate to stars with convective envelopes. 

Both components of the DG Mic system are too massive to fit that scenario. An earlier episode of mass transfer from the primary to the secondary could be a possible explanation for the anomaly. Given that the components occupy approximately 50\% of their respective Roche lobes, future episodes of mass transfer might occur. Additional spectroscopy of the system, at a substantially higher resolution than we were able to obtain in this study, might provide greater clarity about the metal content of the components and allow determination of the physical parameters with greater accuracy. This might resolve the observed radius anomaly to some extent. \citet{Bulut_2021} found a similar radius-age discrepancy between the two components of another eclipsing binary system (EPIC 216075815) with component masses almost identical to those of DG Mic. They refer to the low resolution of the spectra on which their analysis was based as a possible factor causing the observed discrepancy. However, our spectroscopy of DG Mic attained significantly higher resolution and our determined errors imply that the mass anomaly observed in the secondary component is real. As a further point of interest, the components of DG Mic have practically identical masses, which qualifies it as a \lq\lq{twin}\rq\rq eclipsing binary system -- as is EPIC 216075815.

\citet{Bulut_2021} note that the evolutionary status of twin stars has been a decades-long subject of debate and that the discovery (and analysis) of a twin eclipsing binary system is a rare event. Our finding of another apparent radius anomaly in a twin eclipsing binary system (with component masses comparable to those of EPIC 216075815) might facilitate an improved understanding of the evolution of close binary systems.


\begin{table}
  \centering
\caption{The absolute parameters of DG Mic.}
\label{table3}
\small
\begin{tabular}{lll}
\hline
Parameter		& Primary		& Secondary		\\
\hline
$A$ (R$_{\odot}$)	& \multicolumn{2}{c}{12.11($\pm$0.21)}  \\
$M$ (M$_{\odot}$)	& 1.65($\pm$0.12)	& 1.64($\pm$0.18)	       \\
$R$ (R$_{\odot}$)	& 1.63($\pm$0.10)	& 1.91($\pm$0.13)	       \\
log $g$ (cgs)		& 4.23($\pm$0.02)	& 4.09($\pm$0.01)	       \\
$T$ (K)			& 7300($\pm$200) 	& 7025($\pm$204)	      \\
$L$ (L$_{\odot}$)	& 6.80($\pm$1.65)	& 8.00($\pm$1.64)	  \\
$M_{bol}$ (mag)	& 2.66($\pm$0.26)	& 2.48($\pm$0.22)	       \\
$M_{V}$ (mag)		& 2.43($\pm$0.26)	& 2.19($\pm$0.22)	      \\
$M_{V}$ $(system)$ (mag)		 &  \multicolumn{2}{c}{1.62($\pm$0.22)}   \\
$d$ (pc)              &     \multicolumn{2}{c}{224($\pm$20)}    \\
$d_{GAIA}$ (pc) DR2   &    \multicolumn{2}{c}{226($\pm$4)\textsuperscript{a}}    \\
$d_{GAIA}$ (pc) DR3 & \multicolumn{2}{c}{221($\pm$1)\textsuperscript{b}}          \\
\hline
$^{a}$ \citet{Gaia_2018},
$^{b}$ \citet{Gaia_2023}. \\
\end{tabular}
\end{table}

\begin{figure}
\centering
\includegraphics[width=130mm]{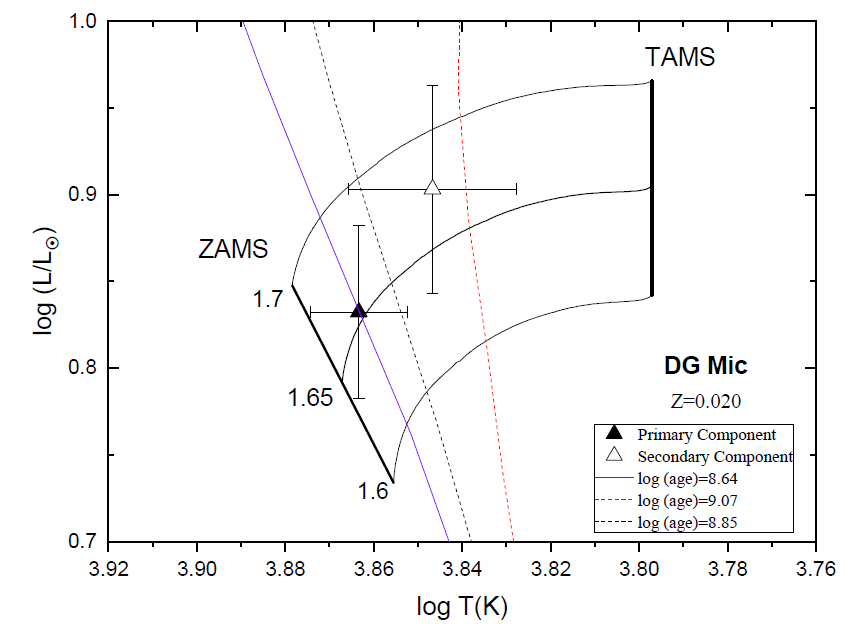} \\
\includegraphics[width=120mm]{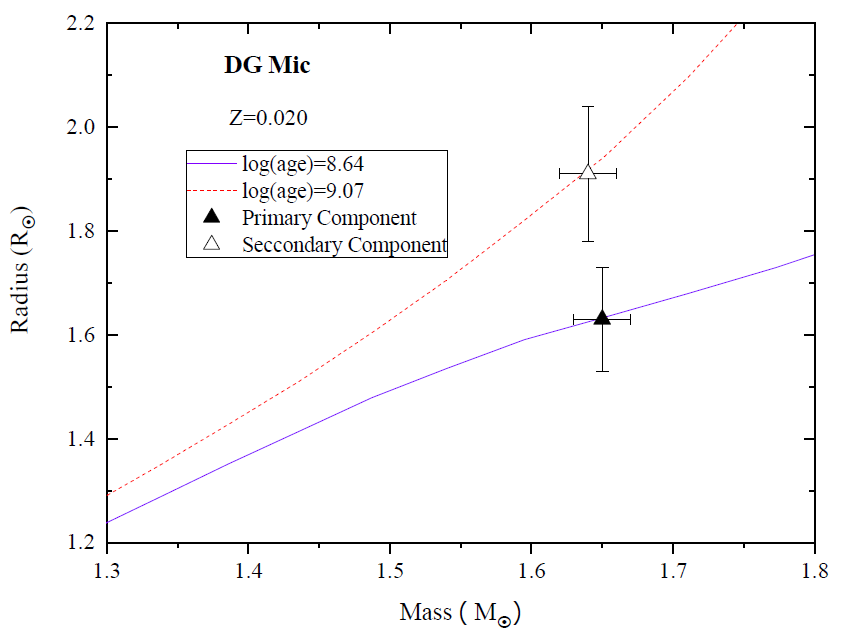}
\caption{Comparison of DG~Mic with Geneva evolutionary models (Mowlavi, et al., 2012) for $Z = 0.020$. 
\textit{Upper panel:} $\log T_{\rm eff}$--$\log L$ diagram. 
\textit{Lower panel:} mass--radius diagram. The best-fitting isochrone that simultaneously reproduces both components (within their uncertainties) corresponds to an age of 713~Myr 
(black dashed line). Isochrones that best match the primary and secondary individually are shown with the violet solid and red dashed lines, respectively. Insets display the corresponding ages of the three isochrones. Filled and open triangles denote the primary (hotter) and secondary (cooler) 
components, with error bars indicating observational uncertainties. The zero-age main sequence (ZAMS) and terminal-age main sequence (TAMS) are shown as thick black lines.}
\label{fig4}
\end{figure}

\section*{Acknowledgements}

We would like to thank an anonymous referee for his/her helpful and critical suggestions greatly improving the overall level of the manuscript. This research was supported by the Scientific Research Projects Directorate of \c{C}anakkale Onsekiz Mart University under the Grant No: FYL-2020-3328. This study was part of the Mücahit Kuz Master's thesis. CAE thanks the South African National Research Foundation and the University of Johannesburg for financial support. All the authors thank the SAAO for observing time. The authors thank the ASAS, SWASP and TESS teams for making all of the observations readily accessible on public platforms.


\appendix
\section{Appendix: Radial velocity values (with errors) of the components of DG Mic}

\scriptsize
\begin{small}
\begin{longtable}{cccccc}
	\caption{Radial velocity values (with errors) of the components of DG~Mic obtained using the cross-correlation method.}%
	\label{TableA1} \\ 
	\hline
	Observation Time  & Phase	 &	$RV_{1}$		& $\sigma_{1}$		& $RV_{2}$	& $\sigma_{2}$	\\
	HJD	    	      &	         & (km/s)			& (km/s)			& (km/s)	& (km/s)	\\
	\hline
	\endfirsthead

\multicolumn{6}{c}%
{\tablename\ \thetable\ -- \textit{Continued from previous page}} \\
\hline
Observation Time			& Phase	&	$RV_{1}$		& $\sigma_{1}$		& $RV_{2}$	& $\sigma_{2}$		\\
HJD	    		&	         & (km/s)			& (km/s)			& (km/s)		& (km/s)			\\
\hline
\endhead

\hline \multicolumn{6}{r}{\textit{Continued on next page}} \\
\endfoot

\hline
\endlastfoot

2459063.3222	& 0.150	& -90.3				& 8.5				& --		& --	\\
2459063.3280	& 0.152	& -80.8				& 8.1				& --		& --	\\
2459063.3329    & 0.154 & -81.0             & 10.7              & --        & --  \\
2459063.5973    & 0.252 & -123.3            & 22.6              & 99.1     & 28.4    \\
2459063.6030    & 0.254 & -126.2            & 21.1              & 86.3     & 29.3    \\
2459063.6089    & 0.256 & --                & --                & 92.0     & 26.0    \\
2459063.6146    & 0.258 & --                & --                & 91.3     & 35.1    \\
2459155.3140    & 0.272 & -100.1            & 14.3              & 98.2     & 29.8    \\
2459128.3552    & 0.272 & -105.5            & 13.3              & --       & --    \\
2459128.3611    & 0.274 & -103.6            & 12.6              & --       & --    \\
2459155.3217    & 0.275 & -105.6            & 16.8              & 107.7    & 30.1    \\
2459128.3669    & 0.277 & -104.1            & 13.9              & --       & --     \\
2459155.3264    & 0.277 & -107.8            & 16.3              & 116.9     & 43.2     \\
2459155.3311    & 0.278 & -112.6            & 15.5              & 112.0     & 39.1     \\
2459128.3909    & 0.286 & -116.2            & 14.6              & 118.3     & 16.3     \\
2459128.3974    & 0.288 & -119.0            & 15.1              & --        & --     \\
2459128.4046    & 0.291 & -116.3            & 15.3              & 114.3     & 17.2     \\
2459128.4116    & 0.293 & -116.7            & 15.8              & 121.5     & 17.6    \\
2459128.4186    & 0.296 & -114.3            & 14.8              & 112.6     & 18.4    \\
2459128.4255    & 0.298 & -109.1            & 16.3              & 102.9     & 18.9    \\
2459061.5589    & 0.495 &  --               & --                & --        & --      \\
2459064.5195    & 0.594 & 36.3              & 15.1              & -67.6     & 22.1    \\
2459064.5264    & 0.596 & 47.3              & 13.1              & --        & --      \\
2459064.5325    & 0.598 & --                & --                & --        & --      \\
2459064.5590    & 0.608 & 63.3              & 15.5              & -80.1     & 21.6    \\
2459064.5649    & 0.610 & 48.7              & 15.6              & -79.8     & 43.0    \\
2459064.5702    & 0.612 & 41.2              & 19.4              & -90.3     & 51.8    \\
2459064.5750    & 0.614 & 46.1              & 17.7              & -81.9     & 31.3    \\
2459156.2959    & 0.636 & 69.0              & 18.1              & -80.8     & 29.9    \\
2459156.3008    & 0.638 & 67.8              & 18.8              & -85.7     & 29.9    \\
2459156.3056    & 0.640 & 80.3              & 20.2              & --        & --      \\
2459129.3516    & 0.642 & 77.7              & 14.4              & -86.1     & 27.7    \\
2459129.3575    & 0.644 & 84.7              & 14.3              & -101.4    & 34.8    \\
2459129.3633    & 0.646 & 92.3              & 14.7              & -93.7     & 31.0     \\
2459129.3690    & 0.648 & 90.9              & 14.7              & -106.3    & 40.5      \\
2459129.3935    & 0.657 & 104.1             & 15.5              & -104.3    & 35.0     \\
2459129.3995    & 0.660 & 109.5             & 13.0              & -91.5     & 35.8     \\
2459129.4053    & 0.662 & 109.2             & 15.3              & -103.0    & 31.1      \\
2459129.4112    & 0.664 & 106.4             & 14.1              & -94.6     & 30.3      \\
2459070.3077    & 0.741 & 113.1             & 14.3              & -118.9    & 28.1      \\
2459070.3144    & 0.743 & 118.1             & 14.9              & -107.0    & 26.9      \\
2459070.3233    & 0.746 & 122.0             & 13.3              & -106.5    & 32.7      \\
2459062.3311    & 0.782 & 111.9             & 25.8              & -106.2    & 27.1      \\
2459062.3841    & 0.802 & 115.8             & 22.4              & -82.9     & 34.0      \\
2459062.3894    & 0.804 & 116.4             & 22.5              & -85.2     & 31.4      \\
2459062.3946    & 0.805 & 117.5             & 26.6              & -80.2     & 48.3      \\
2459062.4924    & 0.842 & --                & --                & -85.6     & 43.2      \\
2459062.4973    & 0.844 & --                & --                & -84.7     & 40.5      \\
2459062.5023    & 0.845 & --                & --                & -88.1     & 38.0      \\
2459062.5069    & 0.847 & --                & --                & -91.0     & 35.8      \\
2459062.5891    & 0.878 & --                & --                & -65.0     & 24.7      \\
2459062.5938    & 0.879 & 41.9              & 21.5              & --       & --        \\
2459062.5986    & 0.881 & 46.8              & 14.7              & --        & --        \\
\hline
\end{longtable}
\end{small}


\end{document}